\documentclass{svjour3}                     
\smartqed
\usepackage{graphicx}
\usepackage{graphics}
\usepackage[centerlast,sc,small]{subfigure}
\usepackage{slashbox}
\usepackage{amsmath}
\usepackage{color}
\usepackage{colortbl}
\usepackage{appendix}
\usepackage{mathptmx}      
\usepackage{txfonts}
\usepackage[authoryear]{natbib}
\usepackage{url}
\def\la{\mathrel{\mathchoice {\vcenter{\offinterlineskip\halign{\hfil
$\displaystyle##$\hfil\cr<\cr\sim\cr}}}
{\vcenter{\offinterlineskip\halign{\hfil$\textstyle##$\hfil\cr
<\cr\sim\cr}}}
{\vcenter{\offinterlineskip\halign{\hfil$\scriptstyle##$\hfil\cr
<\cr\sim\cr}}}
{\vcenter{\offinterlineskip\halign{\hfil$\scriptscriptstyle##$\hfil\cr
<\cr\sim\cr}}}}}

\begin{document}

\title{Numerical integration of dynamical systems with Lie series}
\subtitle{relativistic acceleration and non-gravitational forces}



\author{D. Bancelin   
         \and D. Hestroffer
         \and W. Thuillot
}


\institute{D. Bancelin \at
              IMCCE, Paris Observatory, CNRS, UPMC\\ 
              77, Av. Denfert-Rochereau 75014 Paris France \\
              Tel.: +33 1 4051 2271\\
              Fax:  +33 1 4633 2834\\
              \email{bancelin@imcce.fr}\\
           \and
           D. Hestroffer \at
              Tel.: +33 1 4051 2260\\
              \email{hestro@imcce.fr} \\ 
           \and
           W. Thuillot \at
              Tel.: +33 1 4051 2262\\
              \email{thuillot@imcce.fr} 
}

\date{Received: date / Accepted: date}
\maketitle

\begin{abstract}
The integration of the equations of motion in  gravitational dynamical systems---either in our Solar System or for extra-solar planetary systems---
being non integrable in the global case, is usually performed by means of numerical integration. Among the different numerical techniques available
for solving ordinary differential equations, the numerical integration using Lie series has shown some advantages. In its original
form \citep{hanslmeier84}, it was limited to the $N-$body problem where only gravitational interactions are taken into account.

We present in this paper a generalisation of the method by deriving an expression of the Lie-terms when other major forces are considered. As a matter
of fact, previous studies had been made but only for objects moving under gravitational attraction. If other perturbations are added, the Lie
integrator has to be re-built. In the present work we consider two cases involving  position and position-velocity dependent perturbations:
relativistic acceleration in the framework of General Relativity and a simplified force for the Yarkovsky effect. A general iteration procedure is
applied to derive the Lie series to any order and precision. We then give an application to the integration of the equation of motions for typical
Near-Earth objects and planet Mercury.

\end{abstract}

\section{Introduction}

The integration of the equations of motion in gravitational dynamical systems, since they are not integrable in closed form with more than 2 massive
bodies involved, is usually performed by means of numerical integration. This can be done over times scales of a few days or a century for
establishing ephemerides, or longer periods of several million years to establish stability properties \citep{pal07}. It concerns bodies in
our own Solar System as well as extra-solar planetary systems. Several algorithms have been used or developed in particular in celestial mechanics to
solve ordinary differential equations (hereafter ODEs) such as Runge-Kutta method, Everhart's integrator, Bulirsch-Stoer, Adams, Lie Series,
etc. Numerical integration by means of Lie series \citep{hanslmeier84}, which is based on generating the Taylor expansion of an ODEs solution, has
shown some interest. It is a high speed symplectic integrator adapted to the case of the $N-$body problem. The construction of Lie-series integrator
was first investigated by \cite{hanslmeier84}, \cite{delva85} and \cite{lichtenegger84}. It has also been developed for studying the stability and
dynamical evolution of Near-Earth Objects (NEOs) in the inner Solar System \citep{dvorak99,pal07}, or the stability of extra-solar systems
\citep{schwarz05}. It has been developed to also integrate the case of varying masses \citep{dvorak83} and the damped oscillator \citep{dvorak83b}. 

In the previous works, a recurrence formula for the Lie-terms has been developed but only for masses moving under their mutual gravitational
attraction. As a consequence, it cannot be used as such for masses (e.g. asteroids and comets) for which relativistic or non-gravitational forces are
non-negligible. 

When dealing with comets, NEOs, highly eccentric orbits, and extra-solar systems, other forces and acceleration can come into play.
Objects close to their star with significant eccentricity will show a precession of their perihelion from the relativistic acceleration. Such
perihelion precession will affect the transit time determination in extra-solar planets quests \citep{pal08}. It can also affect general studies of
long-term stability as it has been put into evidence in our Solar System \citep{laskar09}, since secular resonances can eventually be avoided or
induced from this additional frequency. It will also affect the ephemerides of comets and NEOs because these can have small perihelion distance and
large eccentricities. In addition to the relativistic acceleration, NEOs can also be affected by the Yarkovsky effect (\cite{vokrouhlicky98},
\cite{bottke02}) . This effect is caused by a recoil force of anisotropically emitted thermal radiation and is of importance for small bodies ($\la
20\,$km in diameter) close to the Sun, for much smaller bodies the solar radiation pressure will have a larger influence. The Yarkovsky effect can be
divided into a seasonal and diurnal effect, depending on the orientation of the spin axis. It is the diurnal force that will have a major effect on
the orbit of an asteroid, by yielding a secular variation of the semi-major axis \citep{bottke02}. 

Although the use of Lie-series for integrating the $N-$body problem has been studied by \cite{grobner67} it was not successfully implemented until the
major breakthrough from \cite{hanslmeier84} who provided a recurrence formula for practical use of the series expansion. As a matter of fact, the Lie
integrator has to be redesigned when other forces are taken into account.

After a brief introduction of the Lie operator and use of Lie series for the numerical integration of the $N-$body problem (see
section~\ref{sec:intro}), we present in section \ref{sec:3} a generalised expression of the Lie Series integration by providing the recurrence formula
applicable to the case of a relativistic acceleration and a velocity-dependent force (applicable to the case of the Yarkovsky effect). We give some
numerical tests and an application to the integration of the equation of motions for NEOs and planet Mercury in section \ref{sec:4}. We then
conclude
and discuss future possible developments in section \ref{sec:conclusion}.

\section{Lie operator and Lie series}
\label{sec:intro}
\subsection{Lie derivative}

A brief sum up of Lie operator will be given in this section, as it has already been studied in \cite{hanslmeier84,dvorak83,eggl2010}. The Lie 
operator $D$ over a manifold of dimension $N$ is defined as:
\begin{equation}
\label{def_Lie}
 D = \sum_{k=1}^N \theta_{k}\frac{\partial \cdot}{\partial z_{k}}
\end{equation}
where $\theta$ = ($\theta_{1}(\vec{z})$, ..., $\theta_{N}(\vec{z})$) is a holomorphic function over a domain $\cal D$ in the $\bf z$-space, that is,it
can be expanded in a converging power series, and 
 $\vec{z}$ = (z$_{1}$, ..., z$_{N}$).\\

Applying $n-$times this operator on a function f($\vec{z}$), holomorphic over the same domain $\cal D$, we have:
\begin{equation}
 D\,f = \sum_{k=1}^N \theta_{k} {\partial f \over \partial z_{k}}
 \quad ; \quad
 D^{n}f = D(D^{n-1}f)
\end{equation}
 We remind here some useful properties of linearity and Leibniz-rule of Lie operator such as:

\begin{eqnarray}
\label{properties}
 \left\{
\begin{array}{l}
 D^{\scriptscriptstyle n}( f + g) = D^{\scriptscriptstyle n}\,f + D^{\scriptscriptstyle n}\,g\\
 D^{\scriptscriptstyle n}( f \cdot g) = \sum_{k=0}^{n}\,\binom{n}{k}\,D^{\scriptscriptstyle k}\,f\,D^{\scriptscriptstyle n-k}\,g
\end{array}
\right.
\end{eqnarray}

The Lie-series are then defined as:
\begin{equation}
 L(\vec{z},t) \equiv \sum_{\mu=0}^{\infty} \frac{t^{\mu}}{\mu!}\,D^{\mu}f(\vec{z})  ~~~\mbox{} \mu \in N
 \end{equation}
which is converging over $\cal D$ \citep{grobner67}. And from its similarity with the expansion of the exponential function, we can write
symbolically:
\begin{equation}
   L(\vec{z},t) \equiv e^{tD}f(\vec{z})
\label{E:}
\end{equation}

This expression can be used to solve a system of ODEs such as:
\begin{equation}
\label{eq_diff}
  \dot{z_k} = \theta_{k}(\vec{z})
\end{equation}
The solution is $z_k = e^{tD}\,(\xi_{k})$, where $\xi_{k}$ are the initial conditions, and $D = \sum_{k=1}^N \theta_{k}(\xi) {\partial \cdot
\over \partial \xi_{k}}$. Similarly, the approximate solution of $\vec{\dot{z}} = \theta(\vec{z})$ at time t+$\Delta$t is:
\begin{equation}
 \vec{z}(t+\Delta t) = L(\vec{z},\Delta t) = e^{\Delta tD}\vec{z}(t)
\end{equation}

\subsection{The $N-$body problem}
We can apply the Lie integration method to solve the equations of motion of the $N-$body problem. Since we will generalise this problem by
including additional forces, we remind here the general steps of the procedure for the purely gravitational $N-$body case as given in
\cite{hanslmeier84}.

According to the law of attraction, the Newtonian acceleration resulting from the forces acting on a particle $\nu$ is:
\begin{equation}
 \label{eq_mvt}
 \ddot{\vec{x}}_{\nu} = G\,\sum_{\scriptscriptstyle \mu=1,\mu\ne \nu}^{\scriptscriptstyle N} M_{\mu}\,\frac{(\vec{x}_{\scriptscriptstyle \mu}
-\vec{x}_{\nu})}{\|\vec{x}_{\scriptscriptstyle \mu} - \vec{x}_{\scriptscriptstyle \nu} \|^{3}}
\end{equation}
where $G$ is the gravitational constant and $\vec{x}_{\scriptscriptstyle \nu}$ the barycentric position vector of particle $\nu$ with mass
$M_{\scriptscriptstyle \nu}$. By introducing a new variable $v$, the 2$^{\scriptscriptstyle nd}$ order system (\ref{eq_mvt}) can be transform
into a system of 1$^{\scriptscriptstyle st}$ order differential equations.
\begin{eqnarray}
  \dot{\vec{x}}_{\scriptscriptstyle \nu} & = & \vec{v}_{\scriptscriptstyle \nu} \label{eq_diff1} \\
  \dot{\vec{v}}_{\scriptscriptstyle \nu} & = & G\,\sum_{\scriptscriptstyle \mu=1,\mu\ne \nu}^{\scriptscriptstyle  N} 
M_{\mu}\,\frac{(\vec{x}_{\scriptscriptstyle \mu} - \vec{x}_{\scriptscriptstyle \nu})}{\|\vec{x}_{\scriptscriptstyle \mu} - 
\vec{x}_{\scriptscriptstyle \nu} \|^{3}}
  =  GM_{\star}\,\sum_{\scriptscriptstyle \mu=1,\mu\ne \nu}^{\scriptscriptstyle N} m_{\scriptscriptstyle \mu}\,\,  {\mathbf x_{\scriptscriptstyle \nu
\mu} \, \rho_{\scriptscriptstyle \nu \mu}^{-3}}
\label{eq_diff2}
\end{eqnarray}
\noindent
such as $\rho_{\scriptscriptstyle \nu \mu}  =  \| \vec{x}_{\scriptscriptstyle \mu} - \vec{x}_{\scriptscriptstyle \nu}\|$
and $\vec{x}_{\scriptscriptstyle \nu \mu} = \vec{x}_{\scriptscriptstyle \mu} - \vec{x}_{\scriptscriptstyle \nu}$.\\
Here M$_{\star}$ is a conversion factor to express the mass $M_{\mu}$ of the perturbing body in the unit of star mass. Its numeric value is
equal to the central star's mass of the planetary system. Thus m$_\mu$ is the mass of the perturbing body in the unit of star mass (or
Solar mass if $GM_\star$= k$^{\scriptscriptstyle 2}$ with k representing the Gauss constant\footnote{$G$ has to be converted first in
AU$^{\scriptscriptstyle 3}$\,kg$^{\scriptscriptstyle {-1}}$\,day$^{\scriptscriptstyle {-2}}$}).

The Lie operator for the Newtonian gravitational $N-$body system has been given by \cite{eggl2010} and can be expressed as :
\begin{equation}
\label{Lie_Oper_Pla}
 D = \sum_{\nu=1}^{N} \left( \vec{v}_{\scriptscriptstyle \nu} \, 
 \frac{\partial \cdot}{\partial \vec{x}_{\scriptscriptstyle \nu }} 
 + GM_{\star}\sum_{\scriptscriptstyle \mu=1,\mu\ne \nu}^{\scriptscriptstyle N} m_{\scriptscriptstyle \mu}\, \vec{x}_{\scriptscriptstyle \nu \mu}\, 
\rho_{\scriptscriptstyle \nu \mu}^{-3}\,\frac{\partial \cdot}{\partial \vec{v}_{\scriptscriptstyle \nu}} \right)
\end{equation}
According to Eq. (\ref{def_Lie}) and (\ref{eq_diff}) the solutions of Eq. (\ref{eq_diff1}) and (\ref{eq_diff2}) are given by the series expansion:
\begin{equation}
 \vec{x}_{\nu}(\tau) = e^{\tau D}\vec{x}_{\nu}(0) = \left(  \sum_{n=0}^{\infty} \frac{\tau^{n} D^{n}}{n!}\right) \vec{x}_{\nu}(0) 
\end{equation}
\begin{equation}
 \vec{v}_{\nu}(\tau) = e^{\tau D}\vec{v}_{\nu}(0) = \left(  \sum_{n=0}^{\infty} \frac{\tau^{n} D^{n}}{n!}\right) \vec{v}_{\nu}(0) 
\end{equation}
where $\tau$ is the current step-size and is defined by:
\begin{eqnarray*}
 \tau = t_{\scriptscriptstyle j} - t_{\scriptscriptstyle j-1}
\end{eqnarray*}

For this method to be of practical use, a recurrence formula has to be given to derive any order of the Lie operator. This was given in
\cite{eggl2010} for the $N-$body problem:\\
\begin{eqnarray}
 D^{\scriptscriptstyle n} \vec{x}_{\scriptscriptstyle \nu} = GM_{\star}\sum_{\scriptscriptstyle \mu = 1, \mu \ne \nu}^{\scriptscriptstyle N} 
m_{\scriptscriptstyle \mu} \sum_{\scriptscriptstyle k = 0}^{\scriptscriptstyle n-2} \binom{n-2}{k} D^{\scriptscriptstyle k} \Phi_{\scriptscriptstyle
\nu \mu} D^{\scriptscriptstyle n-2-k} \vec{x}_{\scriptscriptstyle \nu \mu}
\end{eqnarray}

where $\Phi_{\scriptscriptstyle \nu \mu}  =  \rho^{\scriptscriptstyle -3}_{\scriptscriptstyle \nu \mu}$.

The evolution of $\Phi_{\scriptscriptstyle \nu \mu}$ is given by:

\begin{eqnarray}
\label{Dphi}
 D^{\scriptscriptstyle n}\, \Phi_{\scriptscriptstyle \nu \mu} = \rho^{\scriptscriptstyle -2}_{\scriptscriptstyle \nu \mu}\, \sum^{\scriptscriptstyle
n-1}_{\scriptscriptstyle k=0} a_{\scriptscriptstyle n,k+1} D^{\scriptscriptstyle n-1-k}\, \Phi_{\scriptscriptstyle \nu \mu}  D^{\scriptscriptstyle
k}\, \Lambda_{\scriptscriptstyle \nu \mu}
\end{eqnarray}

the \textit{a$_{\scriptscriptstyle i,j}$} coefficients being defined for $n\ge\,1$:

\begin{eqnarray*}
 \left\{
 \begin{array}{l}
 a_{\scriptscriptstyle n,n}  =  -3 \\
 a_{\scriptscriptstyle n,k}  =  a_{\scriptscriptstyle n-1,k-1} + a_{\scriptscriptstyle n-1,k}	 ~~~~\mbox{for} ~~\mbox k>1 \\
 a_{\scriptscriptstyle n,1}  =  a_{\scriptscriptstyle n-1,1} - 2 
 \end{array}
 \right.
\end{eqnarray*}

$\Lambda_{\scriptscriptstyle \nu \mu}$ is defined by: $\Lambda_{\scriptscriptstyle \nu \mu} = \vec{x}_{\scriptscriptstyle \nu \mu} \cdot 
\vec{v}_{\scriptscriptstyle \nu \mu}$
and its evolution is ruled by:

\begin{eqnarray}\label{E:lambda}
\label{dlambda}
 D^{\scriptscriptstyle n}\, \Lambda_{\scriptscriptstyle \nu \mu} = \sum^{\scriptscriptstyle n}_{\scriptscriptstyle k=0} \binom{n}{k} 
D^{\scriptscriptstyle n-k}\, \vec{x}_{\scriptscriptstyle \nu \mu} D^{\scriptscriptstyle k}\, \vec{v}_{\scriptscriptstyle \nu \mu}
\end{eqnarray}

where

\begin{eqnarray}
 D^{\scriptscriptstyle n}\, \vec{v}_{\scriptscriptstyle \nu \mu} = D^{\scriptscriptstyle n+1}\, \vec{x}_{\scriptscriptstyle \nu \mu} 
\end{eqnarray}

Now if additionnal forces or accelerations act on the particule $\nu$, we have to re-write the system (\ref{eq_diff1}), (\ref{eq_diff2}), the Lie
operator (\ref{Lie_Oper_Pla}), and derive a new recurrence formula to obtain the $n^{th}$ derivative $D^n$ of the Lie operator.

\section{A generalisation of the Lie operator}
\label{sec:3}

This section is dedicated to a generalised expression of the Lie operator for position-dependent and velocity-dependent forces
acting on a particle $\nu$. Let $\vec{H}_{\scriptscriptstyle \nu}$ be the contribution of all the accelerations derived from those forces acting on
body $\nu$.

As in the previous section, the second order ODE is split into a system of six first-order differential equations:
\begin{eqnarray}
  \dot{\vec{x}}_{\scriptscriptstyle \nu}  & =  & \vec{v}_{\scriptscriptstyle \nu}  \\
  \dot{\vec{v}}_{\scriptscriptstyle \nu} & = & \vec{H}_{\scriptscriptstyle \nu} 
\end{eqnarray}
The Lie operator becomes:
\begin{equation}
 D = \sum_{\scriptscriptstyle \mu=1}^{\scriptscriptstyle N} \left[ \vec{v}_{\scriptscriptstyle\mu} \cdot \frac{\partial\cdot}{\partial 
\vec{x}_{\scriptscriptstyle\mu}} +  \vec{H}_{\scriptscriptstyle \mu} \cdot \frac{\partial\cdot}{\partial \vec{v}_{\scriptscriptstyle\mu}} \right] 
\end{equation}
Applying D to $\vec{x}$ , the construction of the Lie terms begins as:
\begin{eqnarray*}
 D^{0}\,\vec{x}_{\scriptscriptstyle \nu} & = & \vec{x}_{\scriptscriptstyle \nu} \\
 D^{1}\,\vec{x}_{\scriptscriptstyle \nu} & = & \vec{v}_{\scriptscriptstyle \nu} \\
 D^{2}\,\vec{x}_{\scriptscriptstyle \nu} & = & \vec{H}_{\scriptscriptstyle \nu}\\
 D^{3}\,\vec{x}_{\scriptscriptstyle \nu} & = & D\,\vec{H}_{\scriptscriptstyle \nu}
\end{eqnarray*}
From which we can deduce the recurrence formula for D$^{n}\vec{x}_{\scriptscriptstyle \nu}$ and  D$^{n}\vec{v}_{\scriptscriptstyle \nu}$ as following:
\begin{eqnarray}
\label{eq-lie_gen}
\left\{
\begin{array}{l}
 D^{\scriptscriptstyle n}\,\vec{x}_{\scriptscriptstyle \nu}   = D^{\scriptscriptstyle n-2}\,\vec{H}{\scriptscriptstyle \nu} ~~~~~~~~~~~~~\text{ n} 
\ge 2 \\
 D^{\scriptscriptstyle n}\,\vec{v}_{\scriptscriptstyle \nu}  =  D^{\scriptscriptstyle n+1}\,\vec{x}_{\scriptscriptstyle \nu} 
\end{array}
\right.
\end{eqnarray}
The next step is to express Eq. (\ref{eq-lie_gen}) according to the expression of the forces which means that we have to find a recurrence formula for
D$^{\scriptscriptstyle n}\, \vec{H}_{\scriptscriptstyle \nu}$.
This will be done in the following sections for three kinds of accelerations: gravitational ($\pmb{\gamma}_{\scriptscriptstyle G}$), relativistic
($\pmb{\gamma}_{\scriptscriptstyle R}$) and non-gravitational accelerations ($\pmb{\gamma}_{\scriptscriptstyle Y}$).
Thus $\vec{H}$ = $\pmb{\gamma}_{\scriptscriptstyle G} + \pmb{\gamma}_{\scriptscriptstyle R} + \pmb{\gamma}_{\scriptscriptstyle Y}$.

\subsection{Lie terms for the gravitational acceleration}

The gravitational acceletation $\vec{\pmb{\gamma}}_{\scriptscriptstyle G/ \nu}$ derived from the gravitational forces of $N$ bodies acting on a body
$\nu$
is:
\begin{eqnarray}
\mathbf{\pmb{\gamma}}_{\scriptscriptstyle G/ \nu} = GM_{\star}\, \sum_{\mu = 1, \mu\ne \nu}^{\scriptscriptstyle N} m_{\scriptscriptstyle \mu}
\Phi_{\scriptscriptstyle \nu \mu,3} \vec{x}_{\scriptscriptstyle \nu \mu} 
\end{eqnarray}
\noindent
Here we have introduced the new variable $\Phi{\scriptscriptstyle \nu\mu,p} = \rho^{\scriptscriptstyle -p}_{\scriptscriptstyle \nu\mu}$ as it will be
useful in the other sections.\\
From the first applications of the Lie operator on $\vec{\pmb{\gamma}}_{\scriptscriptstyle G/ \nu}$:

\begin{eqnarray}
 D\,\vec{\pmb{\gamma}}_{\scriptscriptstyle G/ \nu} & = & GM_{\star}\, \sum_{\mu = 1, \mu\ne \nu}^{\scriptscriptstyle N} m_{\scriptscriptstyle \mu}
\left( 
\Phi_{\scriptscriptstyle \nu \mu,3}\, D\,\vec{x}_{\scriptscriptstyle \nu \mu} + D\, \Phi_{\scriptscriptstyle \nu \mu,3}\, \vec{x}_{\scriptscriptstyle
\nu \mu} 
\right)\\
D^{\scriptscriptstyle 2}\,\vec{\pmb{\gamma}}_{\scriptscriptstyle G/ \nu} & = & GM_{\star}\, \sum_{\mu = 1, \mu\ne \nu}^{\scriptscriptstyle N}
m_{\scriptscriptstyle \mu} \left(
\Phi_{\scriptscriptstyle \nu \mu,3}\, D^{2}\,\vec{x}_{\scriptscriptstyle \nu \mu} + 2\,D\, \Phi_{\scriptscriptstyle \nu \mu,3}\,
\vec{x}_{\scriptscriptstyle \nu \mu} + D^{\scriptscriptstyle 2}\, \Phi_{\scriptscriptstyle \nu \mu,3}\, \vec{x}_{\scriptscriptstyle \nu \mu}
\right)
\end{eqnarray}
we can deduce the evolution of D$^{\scriptscriptstyle n} \,\vec{\pmb{\gamma}}_{\scriptscriptstyle G/ \nu}$:

\begin{eqnarray}
 D^{\scriptscriptstyle n}\,\vec{\pmb{\gamma}}_{\scriptscriptstyle G/ \nu} = GM_{\star}\, \sum_{\mu = 1, \mu\ne \nu}^{\scriptscriptstyle N}
m_{\scriptscriptstyle \mu} \sum_{\scriptscriptstyle k=0}^{\scriptscriptstyle n}\binom{n}{k} D^{\scriptscriptstyle k}\, \Phi_{\scriptscriptstyle \nu
\mu ,3}\, D^{\scriptscriptstyle n-k}\,\vec{x}_{\scriptscriptstyle \nu \mu}
\end{eqnarray}
The evolution of the D$^{\scriptscriptstyle n} \, \Phi_{\scriptscriptstyle \nu \mu, p}$ is ruled by:

\begin{eqnarray}
\label{Dnphi}
 D^{\scriptscriptstyle n}\, \Phi_{\scriptscriptstyle \nu \mu, p} = \rho^{\scriptscriptstyle -2}_{\scriptscriptstyle \nu \mu}\,
\sum^{\scriptscriptstyle n-1}_{\scriptscriptstyle k=0} a_{\scriptscriptstyle n,k+1} D^{\scriptscriptstyle n-1-k}\, \Phi_{\scriptscriptstyle \nu \mu,
p}  D^{\scriptscriptstyle k}\, \Lambda_{\scriptscriptstyle \nu \mu}
\end{eqnarray}
with

\begin{eqnarray*}
 \left\{
 \begin{array}{l}
 a_{\scriptscriptstyle n,n}  =  -p \\
 a_{\scriptscriptstyle n,k}  =  a_{\scriptscriptstyle n-1,k-1} + a_{\scriptscriptstyle n-1,k} ~~~~\mbox{for} ~~\mbox k>1 \\
 a_{\scriptscriptstyle n,1}  =  a_{\scriptscriptstyle n-1,1} - 2 
 \end{array}
\right.
\end{eqnarray*}
and the algorithms for D$^{\scriptscriptstyle n}\,\vec{x}_{\scriptscriptstyle \nu\mu}$ and D$^{\scriptscriptstyle
n}\,\Lambda_{\scriptscriptstyle \nu\mu}$ are given respectively by Eq. (\ref{eq-lie_gen}) and (\ref{E:lambda}).

\subsection{Lie terms for relativistic acceleration}

Because of their complexity, the integration of the EIH (Einstein-Infeld-Hoffman) equations is very slow even using modern computers and are not
suitable for long time integration. Following \cite{beutler05}, the expression of the relativistic acceleration
$\pmb{\gamma}_{\scriptscriptstyle R/ \nu}$ used to generate the Lie-terms is a lighter version (provided that the mass
m$_{\scriptscriptstyle \nu}$ of the massless body is negligle compared to the mass of the central star expressed as:

\begin{eqnarray}
 \label{eq-relat}
 \pmb{\gamma}_{\scriptscriptstyle R/ \nu}  =  \frac{GM_{\star}}{c^2}\,\Phi_{\scriptscriptstyle \mu \nu,3}\left[ \left(
4\,GM_{\star}\,\Phi_{\scriptscriptstyle \mu \nu,1} 
-\vec{\dot r}^{2} \right)\,\vec{r} + 4\,(\vec{r}\cdot \vec{\dot r})\,\vec{\dot r} \right]
\end{eqnarray}
where $c$ represents the speed of light and $\mu$ the Sun.\\
Here, $\vec{r}$ and $\vec{\dot r}$ are the heliocentric position and velocity of the body $\nu$ and are related with the barycentric coordinates by:

\begin{eqnarray*} 
\left\{
\begin{array}{l}
\vec{r} = \vec{x_{\scriptscriptstyle \nu} - x_{\scriptscriptstyle \mu}} = \vec{x}_{\scriptscriptstyle{\mu\,\nu}} \\
\vec{\dot{r}} = \vec{v_{\scriptscriptstyle \nu} - v_{\scriptscriptstyle \mu}} = \vec{v}_{\scriptscriptstyle{\mu\,\nu}} \\
r = \rho_{\scriptscriptstyle \mu\,\nu} 
\end{array}
\right.
\end{eqnarray*}
Equation (\ref{eq-relat}) can be written as follows:

\begin{eqnarray}
 \pmb{\gamma}_{\scriptscriptstyle R/ \nu} = \frac{GM_{\star}}{c^2}\,\Phi_{\scriptscriptstyle \mu\nu,3}\left( \pmb{\gamma}_{\scriptscriptstyle 1}
+
\pmb{\gamma}_{\scriptscriptstyle 2} \right)
\end{eqnarray}
where

\begin{eqnarray}
\left\{
\begin{array}{rl}
 \pmb{\gamma}_{\scriptscriptstyle 1} & =  \vec{x}_{\scriptscriptstyle \mu\nu} \left(4\,GM_{\star}\,\Phi_{\scriptscriptstyle \mu \nu,1}
-\vec{v}_{\scriptscriptstyle \mu\nu}^{2} \right) \\
 \pmb{\gamma}_{\scriptscriptstyle 2} & = 4\,\vec{v}_{\scriptscriptstyle \mu\nu}\,\Lambda_{\scriptscriptstyle \mu\nu}
 \end{array}
\right.
\end{eqnarray}

In the same way, the first derivatives of D$^{\scriptscriptstyle n}\pmb{\gamma}_{\scriptscriptstyle R/ \nu}$ lead to the algorithm:
\begin{eqnarray}
 D^{\scriptscriptstyle n}\,\pmb{\gamma}_{\scriptscriptstyle R/ \nu} = \frac{GM_{\star}}{c^2} \sum_{\scriptscriptstyle k=0}^{n}
\binom{n}{k}\,D^{\scriptscriptstyle k}\,\Phi_{\scriptscriptstyle \mu\nu,3}\, D^{\scriptscriptstyle n-k} \left( \pmb{\gamma}_{\scriptscriptstyle 1} +
\pmb{\gamma}_{\scriptscriptstyle 2} \right)
\end{eqnarray}


We will now express the evolution of the D$^{\scriptscriptstyle n} \pmb{\gamma}_{\scriptscriptstyle i}$ by watching the evolution of their first
derivatives. Using the Lie properties expressed in Eq. (\ref{properties}) 
there is no difficulty to find this evolution. This approach leads to the recurrence formulas:

\begin{eqnarray}
 \left\{
 \begin{array}{l}
   D^{\scriptscriptstyle n}\,\pmb{\gamma}_{\scriptscriptstyle 1} = \sum_{\scriptscriptstyle k=0}^{\scriptscriptstyle n}
\binom{n}{k}\,D^{\scriptscriptstyle
n-k}\,\vec{x}_{\scriptscriptstyle \mu\nu} \left( 4\,GM_{\star}\,D^{\scriptscriptstyle k}\,\Phi_{\scriptscriptstyle \mu\nu,1} - 
 \sum_{k^{\prime} = 0}^{k}\binom{k}{k^{\prime}}\,D^{\scriptscriptstyle k^{\prime}}\,\vec{v}_{\scriptscriptstyle \mu\nu}\,D^{\scriptscriptstyle
k-k^{\prime}}\,\vec{v}_{\scriptscriptstyle \mu\nu} \right) \\
 \vspace{0.1cm}  
  D^{\scriptscriptstyle n}\,\pmb{\gamma}_{\scriptscriptstyle 2} = 4\,\sum_{k=0}^{n}\binom{n}{k}\,D^{\scriptscriptstyle k}\vec{v}_{\scriptscriptstyle
\mu\nu}\,D^{\scriptscriptstyle n-k}\Lambda_{\scriptscriptstyle \mu\nu}
 \end{array}
\right.
\end{eqnarray}

We remind that the evolution of the D$^{\scriptscriptstyle n} \vec{x}_{\scriptscriptstyle \nu}$ and D$^{\scriptscriptstyle n}
\vec{v}_{\scriptscriptstyle \nu}$ 
are given by Eq. (\ref{eq-lie_gen}) and the evolution of the D$^{\scriptscriptstyle n}\,\Phi_{\scriptscriptstyle \nu\mu,p}$ and D$^{\scriptscriptstyle
n}\,\Lambda_{\scriptscriptstyle \nu\mu}$
are given by Eq. (\ref{Dnphi}) and Eq. (\ref{dlambda}).

\subsection{Lie terms for Yarkovsky acceleration}

Yarkovsky force is a non-gravitational perturbation caused by a recoil force of an anisotropically emitted thermal radiation and acting on objects 
 close to the Sun and having diameter smaller than $\approx $20km \citep{bottke02}. This force depends on the physical parameters of the asteroid
(diameter, spin period, spin obliquity, surface density, thermal conductivity, etc...),
and for many asteroids most of these physical parameters can be unknown. While the radial component only affects the orbital velocity, the
transversal 
component causes a change of semimajor axis. Thus the Yarkovsky effect is the strongest non-gravitational force acting on small bodies when comparing 
with the intensity of Poynting-Robertson and solar radiation pressure effect.\\  
This force can be decomposed into two effects \citep{vokrouhlicky99}:
\begin{itemize}
\item the diurnal acceleration that depends on the rotation frequency of the body around its spin axis, is maximum when the spin vector is
perpendicular 
to the orbital plane (it means that obliquity is equal to  0$^\circ$ or 180$^\circ$) and causes a drift in 
semimajor axis (positive or negative depending on the value of the spin obliquity).
\item the seasonal acceleration that depends on the mean motion frequency of the object around the Sun, acts when the spin is in-plane (with obliquity
equals to 90$^\circ$) and 
 causes a steady decrease of the semimajor axis.

\end{itemize}

Many tests have been done to see the impact of Yarkovsky effect on the orbital motion of asteroids and potentially hazardous asteroids (PHAs)
(\cite{giorgini08}, \cite{chesley06}). Those authors show that the Yarkovsky force cannot be neglected as it has an effect on the post-close encounter
orbit of PHAs. Although some physical parameters can be unknown, it is possible to take this effect into account without any assumptions regarding
physical characteristics of asteroids.\\

Following \cite{marsden76}, the non-gravitational acceleration acting on comets can be decomposed into three components: radial, transverse 
and normal. For asteroids, the main non-gravitational perturbation due to the Yarkovsky effect is caused by the transverse component. So, we can 
simply express the acceleration like:

\begin{eqnarray}
 \pmb{\gamma}_{\scriptscriptstyle Y} = A_{\scriptscriptstyle 2}\,g(r)\,\vec{T}
\label{eq_yarko}
\end{eqnarray}

\noindent 
where $\vec{T} = (r\,\vec{\dot r}-(\vec{\dot r}\cdot\vec{r})\,\vec{r}/r)/h$ is the transverse unit vector (in the direction of motion). Here
($\vec{r},\vec{\dot{r}}$) are the heliocentric position 
and velocity of the asteroid and $h = \|\vec{r} \wedge \vec{\dot r}\|$ is the norm of the angular momentum of the asteroid. The coefficient
A$_{\scriptscriptstyle 2}$ is a non gravitational parameter depending on the body. 
This parameter is generally accurately determined with three apparitions of a comet, or with optical and radar observations for asteroids. For
instance, the value 
of A$_{\scriptscriptstyle 2}$ for an asteroid is $\approx 10^{\scriptscriptstyle -14}$ AU/day$^{\scriptscriptstyle 2}$ and $\approx
10^{\scriptscriptstyle -8}$ AU/day$^{\scriptscriptstyle 2}$ for comets.\\
Finally, g(r) is a function depending on the heliocentric distance of the object. For Near-Earth Objects, we can simply express this function as: g(r)
= $\left( \frac{1\,A.U.}{r} \right)^{\scriptscriptstyle 2}$

Using the barycentric coordinates:
\begin{eqnarray*} 
\left\{
\begin{array}{l}
\vec{T}  =  \left( \rho_{\scriptscriptstyle \mu\nu}\,\vec{x}_{\scriptscriptstyle \mu\,\nu} -(\vec{v}_{\mu\,\nu}\cdot\,\vec{x}_{\scriptscriptstyle
\mu\,\nu})\,\vec{x}_{\scriptscriptstyle \mu\,\nu}\,\rho_{\scriptscriptstyle \mu\,\nu}^{\scriptscriptstyle -1} \right)/h\\
 h =  \|\vec{x}_{\scriptscriptstyle \mu\,\nu} \wedge \vec{v}_{\scriptscriptstyle \mu\,\nu}\| 
\end{array}
\right.
\end{eqnarray*}

To express the Lie terms for the Yarkovsky acceleration, it is better to write $\pmb{\gamma}_{\scriptscriptstyle Y}$ like:

\begin{eqnarray}
 \pmb{\gamma}_{\scriptscriptstyle Y} = A_{\scriptscriptstyle 2}\,\Phi_{\scriptscriptstyle \mu\,\nu,2}\,J\, \left (\vec{T_{\scriptscriptstyle 1}} -
\vec{T_{\scriptscriptstyle 2}}\right)
\end{eqnarray}

\noindent
with:

\begin{eqnarray*}
 \left\{
\begin{array}{l}
   \vec{T_{\scriptscriptstyle 1}} = \Phi_{\scriptscriptstyle \mu\,\nu,-1}\,\vec{v}_{\scriptscriptstyle \mu\,\nu} \\
   \vec{T_{\scriptscriptstyle 2}} = \Lambda_{\scriptscriptstyle \mu\,\nu}\,\Phi_{\scriptscriptstyle \mu\,\nu,1}\,\vec{x}_{\scriptscriptstyle \mu\,\nu}
\\
    J = h^{\scriptscriptstyle -1}\\
   \Phi_{\scriptscriptstyle \mu\nu,2} = g(r)
 \end{array}
\right.
\end{eqnarray*}

As in the previous sections, it is easy to find the recurrence formula for the evolution of $\pmb{\gamma}_{\scriptscriptstyle Y}$:

\begin{eqnarray}
 D^{\scriptscriptstyle n} \pmb{\gamma}_{\scriptscriptstyle Y} = A_{\scriptscriptstyle 2}\,\sum_{\scriptscriptstyle k=0}^{\scriptscriptstyle
n}\binom{n}{k}\,D^{\scriptscriptstyle n-k}\left( \vec{T_{\scriptscriptstyle 1}} - \vec{T_{\scriptscriptstyle 2}} \right)\,\sum_{\scriptscriptstyle
k^\prime = 0}^{\scriptscriptstyle k}\,\binom{k}{k^\prime}D^{\scriptscriptstyle k^\prime}\,J\,D^{\scriptscriptstyle
k-k^\prime}\,\Phi_{\scriptscriptstyle \mu\nu,2}            
\end{eqnarray}

\noindent
Let J = $(\vec{h}\cdot\vec{h})^{\scriptscriptstyle {-1/2}}$. Thus, $D\,\textit{J} = -(\vec{h}\cdot\vec{h})^{\scriptscriptstyle
{-3/2}}\,D\,\vec{h}\cdot\vec{h}$. 
If $\Gamma = D\,\vec{h}\cdot\vec{h}$ then 
\begin{eqnarray*}
 D\,J = -J^{\scriptscriptstyle 3}\,\Gamma
\end{eqnarray*}

\noindent
The construction of $D^{\scriptscriptstyle n}\,J$ goes as:

\begin{eqnarray*}
 D^{\scriptscriptstyle 2}\,J & = & J^{\scriptscriptstyle 2}\left( -3\Gamma\,D\,J - J\,D\Gamma \right)\\
 D^{\scriptscriptstyle 3}\,J & = & J^{\scriptscriptstyle 2}\left( -5\Gamma\,D^{\scriptscriptstyle 2}\,J -4\,D\Gamma\,D\,J- J\,D^{\scriptscriptstyle
2}\Gamma \right)
\end{eqnarray*}

\noindent
and we can deduce that:
\begin{eqnarray}
 D^{\scriptscriptstyle n}\,J = J^{\scriptscriptstyle 2}\,\sum_{k=0}^{n-1}\,a_{\scriptscriptstyle {n,k+1}}\, D^{\scriptscriptstyle
n-1-k}\,J\,D^{\scriptscriptstyle k}\Gamma
\end{eqnarray}

\noindent
where the coefficients a$_{\scriptscriptstyle i,j}$ are defined, for n$\ge$1, by:
\begin{eqnarray*}
\left\{
\begin{array}{l}
 a_{\scriptscriptstyle n,n}  =  -1 \\
 a_{\scriptscriptstyle n,k}  =  a_{\scriptscriptstyle n-1,k-1} + a_{\scriptscriptstyle n-1,k} ~~~~\mbox{for} ~~\mbox k>1 \\
 a_{\scriptscriptstyle n,1}  =  a_{\scriptscriptstyle n-1,1} - 2 
\end{array}
\right.
\end{eqnarray*}

\noindent
Using the Leibniz-rule, one can easily note that:
\begin{eqnarray*}
 D(\vec{x}\wedge\vec{v}) = D\,\vec{x}\wedge\vec{v} + \vec{x}\wedge\,D\vec{v}
\end{eqnarray*}

\noindent
Thus
\begin{eqnarray}
\left\{
\begin{array}{l}
\vspace{0.1cm}
 D^{\scriptscriptstyle n}\vec{h} = \sum_{k=0}^{n}\binom{n}{k}\,D^{\scriptscriptstyle k}\,\vec{x}_{\scriptscriptstyle
\mu\,\nu}\wedge\,D^{\scriptscriptstyle n-k}\,\vec{v}_{\scriptscriptstyle \mu\,\nu}\\
 D^{\scriptscriptstyle n}\Gamma = \sum_{k=0}^{n}\binom{n}{k}\,D^{\scriptscriptstyle k+1}\,\vec{h}\cdot D^{\scriptscriptstyle n-k}\,\vec{h}
\end{array}
\right.
\end{eqnarray}

\noindent
Finally, the expression of D$^{\scriptscriptstyle n}\vec{T}_{\scriptscriptstyle 1}$ and D$^{\scriptscriptstyle n}\vec{T}_{\scriptscriptstyle 2}$ can
readily be found:
\begin{eqnarray}
D^{\scriptscriptstyle n}\,\vec{T}_{\scriptscriptstyle 1} = \sum_{k=0}^{n}\binom{n}{k}\,D^{\scriptscriptstyle k}\Phi_{\scriptscriptstyle
\mu\,\nu,-1}\,,D^{\scriptscriptstyle n-k}\,\vec{v}_{\mu\nu}
\end{eqnarray}

\noindent
and
\begin{eqnarray}
D^{\scriptscriptstyle n}\,\vec{T}_{\scriptscriptstyle 2} = \sum_{k=0}^{n}\binom{n}{k}\,D^{\scriptscriptstyle
n-k}\vec{x}_{\mu\nu}\,\sum_{k^{\prime}=0}^{k}\binom{k}{k^{\prime}}\,D^{\scriptscriptstyle k^{\prime}}\Lambda_{\mu\nu}\,D^{\scriptscriptstyle
k-k^{\prime}}\Phi_{\mu\nu,1}
\end{eqnarray}

The evolution of $D^{\scriptscriptstyle n } \vec{x}_{\scriptscriptstyle \nu}$ and $D^{\scriptscriptstyle n } \vec{v}_{\scriptscriptstyle \nu}$ are
again given by 
Eq. (~\ref{eq-lie_gen}).

\section{Numerical tests}
\label{sec:4}

In this section, we provide some numerical tests of validation for the redesigned Lie series. 
Previous CPU and accuracy tests have already been provided in \cite{eggl2010} wherein the authors compared two packages of integrators among them
Radau 15, Bulirsch-Stoer and Lie.
As a complement to the previous CPU tests, we tested the time computation for those integrators when considering massless bodies simultaneously
integrated.
Then, we compare the close encounter results between the Earth and asteroid (99942) Apophis with the Radau and Bulirsh-Stoer integrators. 
Finally, we propose some tests for the relativistic algorithm (towards computation of the precession of Mercury's perihelion) and for the Yarkovsky
effect algorithm (towards computation of secular drift of semi-major axis for some asteroids).\\

\subsection{CPU tests}

As a complement to the previous CPU tests already done \citep{eggl2010}, we propose in this part to investigate the integration time for a 
complete Solar System (Sun to Neptune, including Moon) and some varying number of massless bodies . Those minor bodies are taken from a 
synthetic population of NEOs moving only under gravitation. The purpose of this test is to compare the CPU time efficiency of Lie
and 
 Radau integrators when integrating simultaneously massless bodies. For this test, the internal accuracy for both integrators is set to 10$^{-13}$. 
The ODE solved for Radau is a Nclass = 1 type and the number of terms used for Lie integrator is 11. Finally, those minor bodies are integrated over 
one century.

\begin{figure}[h!]
 \centering{
  \includegraphics[height=4cm]{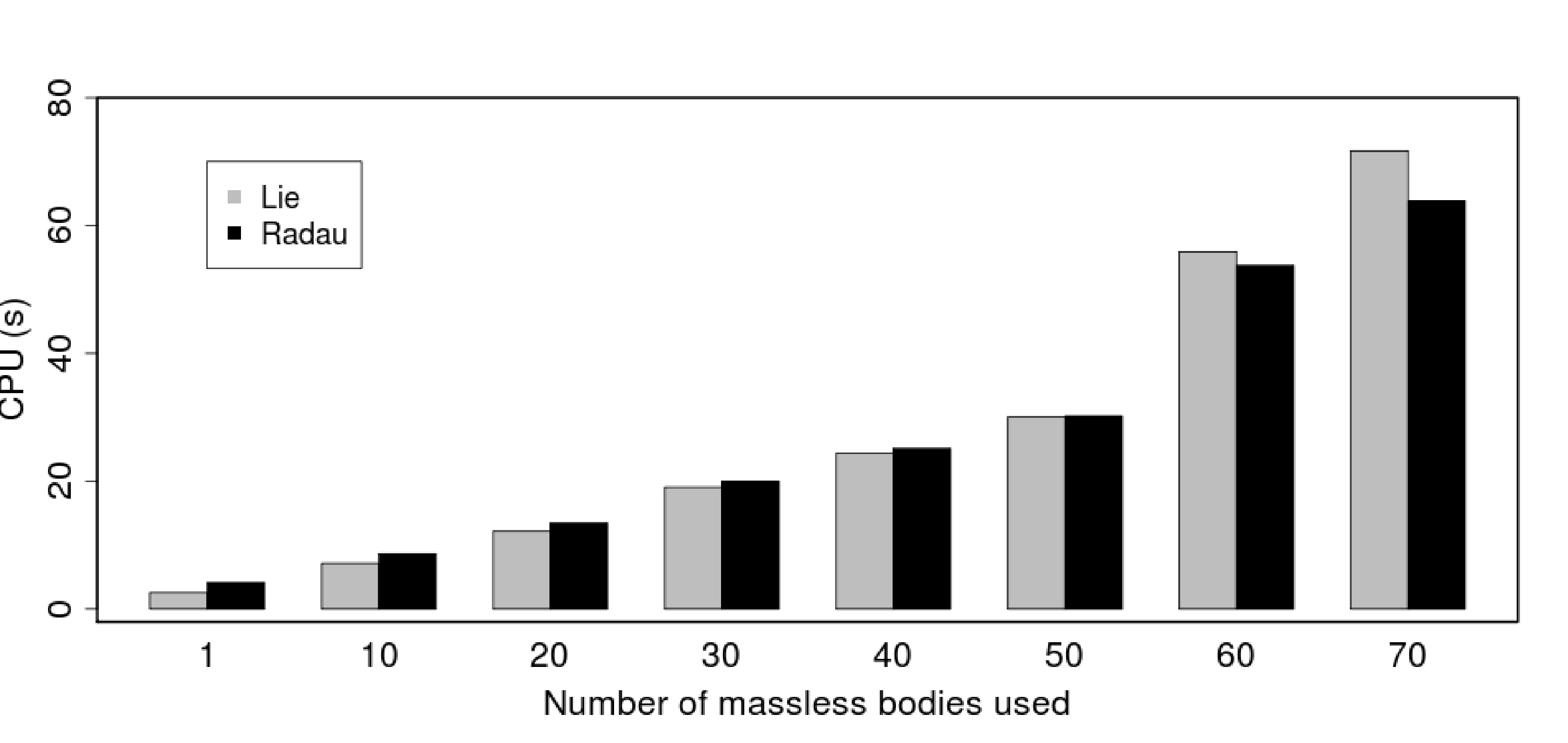}}
\caption{CPU time for Lie and Radau integrators. The abscissa represents the number of minor bodies simultaneously integrated.}
\label{test_cpu}
\end{figure}

We represented in Fig. \ref{test_cpu} the CPU time for both integrators. The abscissa represents the number of massless bodies integrated over 100
years. One can see that the time computation for Lie increases faster than Radau as a function of the number of massless bodies simulatenously
integrated. As a consequence, if Lie is faster for a number less than 50 minor bodies, up to this number, Radau becomes more efficient. More
investigations have to be done to explain this behaviour e.g. requiered work-steps for both integrators, use of the L2 cache instead of L1 cache,
etc, .... So, as noted in \cite{eggl2010} the use of Lie integrator depends on the problem to be solved.

\subsection{Integration of asteroid (99942) Apophis equation of motion}

Asteroid Apophis (previously designed 2004 MN4) is a PHA discovered in 2004. This asteroid was reveled to be a
threatening object since it has a potential impact with Earth in 2036. Because of a deep close encounter in 2029 with Earth, this asteroid is put on a
quasi-chaotic orbit and on possible impact trajectories. 
We propose here to show how Lie integrator can handle the determination of such close approaches distances with an adapted step-size. We compare our
results with Radau and Bulirsh-Stoer integrators. Here, the same dynamical model and initial conditions were used for all three integrators.\\ 

\begin{table}[h!]
 \begin{center}
  \caption{Date and distance of the 2029-close approach of asteroid Apophis computed with three numerical integrators: Radau, Bulirsch and Lie.}
  \label{close-app}
  \begin{tabular}{|c|c|c|}
   \hline
   \hline
   Integrator  & Date of close approach & Distance of close approach (AU) \\
  \hline
   Lie     & 2029  04 13.90716 & 0.000253446941\\
   Bulirsh-Stoer & 2029  04 13.90714 & 0.000253446071\\
   Radau   & 2029  04 13.90725 & 0.000253444524\\
\hline
\hline
  \end{tabular}
 \end{center}
\end{table}

Table \ref{close-app} shows the results for the 2029-close approach of asteroid Apophis computed with three different integrators: Lie, Bulirsh and
Radau. 
We can see that the value of the distance and date of close approach computed with Lie integrator is in a good agreement with the value computed with 
Radau and Burlirsh-Stoer integrator. This test shows that now Lie integrator can handle deep close approach determination with a good precision, 
comparable to other integrators. It could thus be used for close-approach analysis and detection of impacts with Earth.

\subsection{Validation test for General Relativity algorithm}

\begin{figure}[h!]
 \centering{
  \includegraphics[height=8cm]{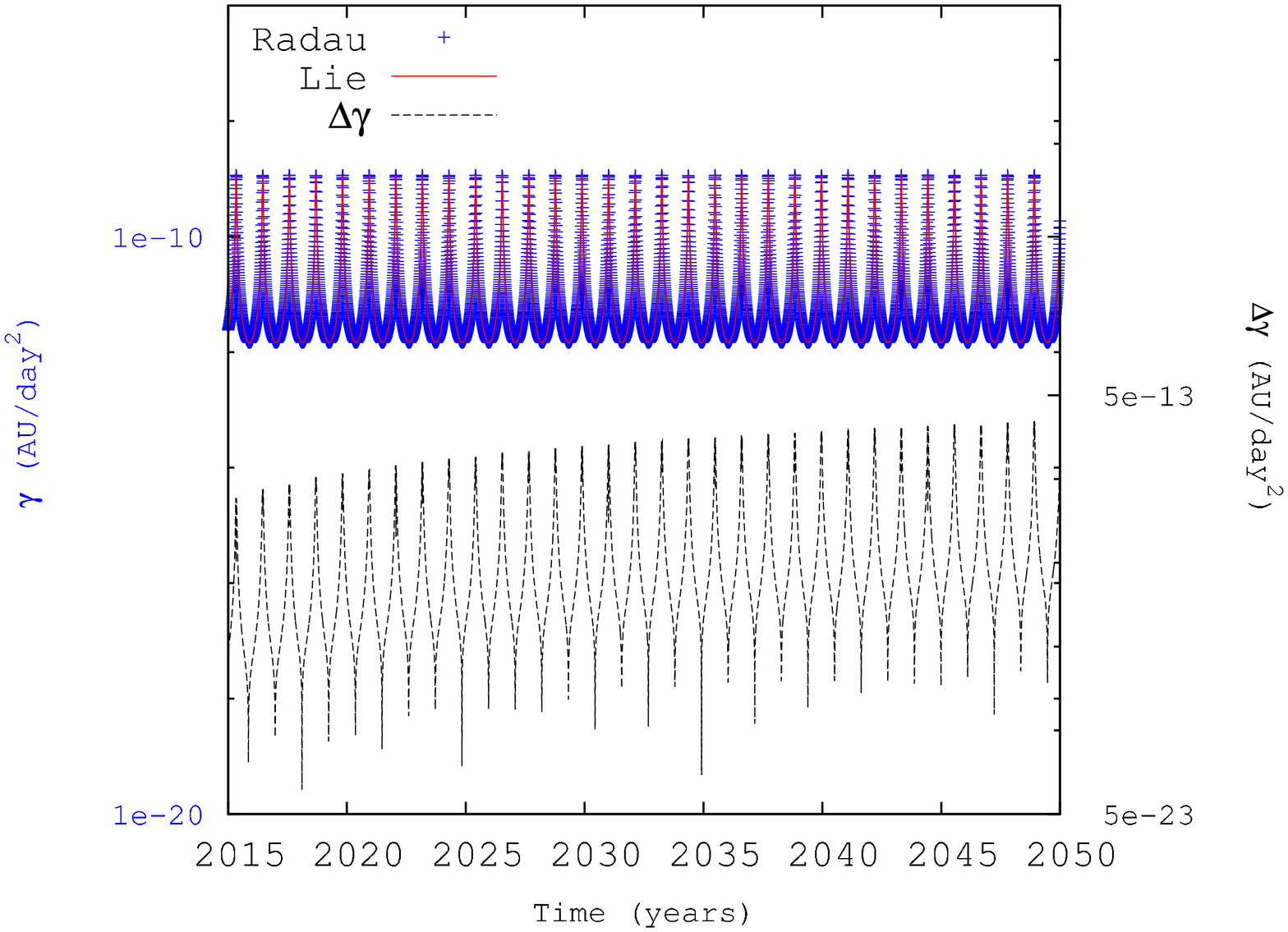}}
\caption{relativistic acceleration calculated with Lie (\textcolor{blue}{+}) and Radau Integrator (\textcolor{red}{---}) and the absolute difference 
$\Delta \pmb{\gamma}$ (- - -) between the value of $\pmb{\gamma}$ calculated with Radau and Lie integrators. The y axis is on a logarithmic scale.}
\label{test_icarus}
\end{figure}

First of all, we tested our general relativity algorithm by integrating the equation of motion of asteroid (1566) Icarus. This asteroid is a good
example for this test in as much as, its 
eccentricity is $\sim$ 0.82 and is known to be sensitive to general relativistic effect. We calculate the relativistic acceleration with Lie series
and then we compare our results with Radau integrator. 
For both integrators, the numerical precision is set to $10^{\scriptscriptstyle -13}$.
Figure \ref{test_icarus} represents the value of this acceleration (between $\sim$ 10$^{\scriptscriptstyle -11}$ and $\sim$ 10$^{\scriptscriptstyle
-9}$)
calculated with Lie and Radau and shows that those value are very close. The absolute difference between those values validates the Lie algorithm as 
this difference lies between $\sim$ 10$^{\scriptscriptstyle -23}$ and $\sim$ 10$^{\scriptscriptstyle -17}$.\\

\begin{figure}[h!]
\centering{
  \includegraphics[height=8cm]{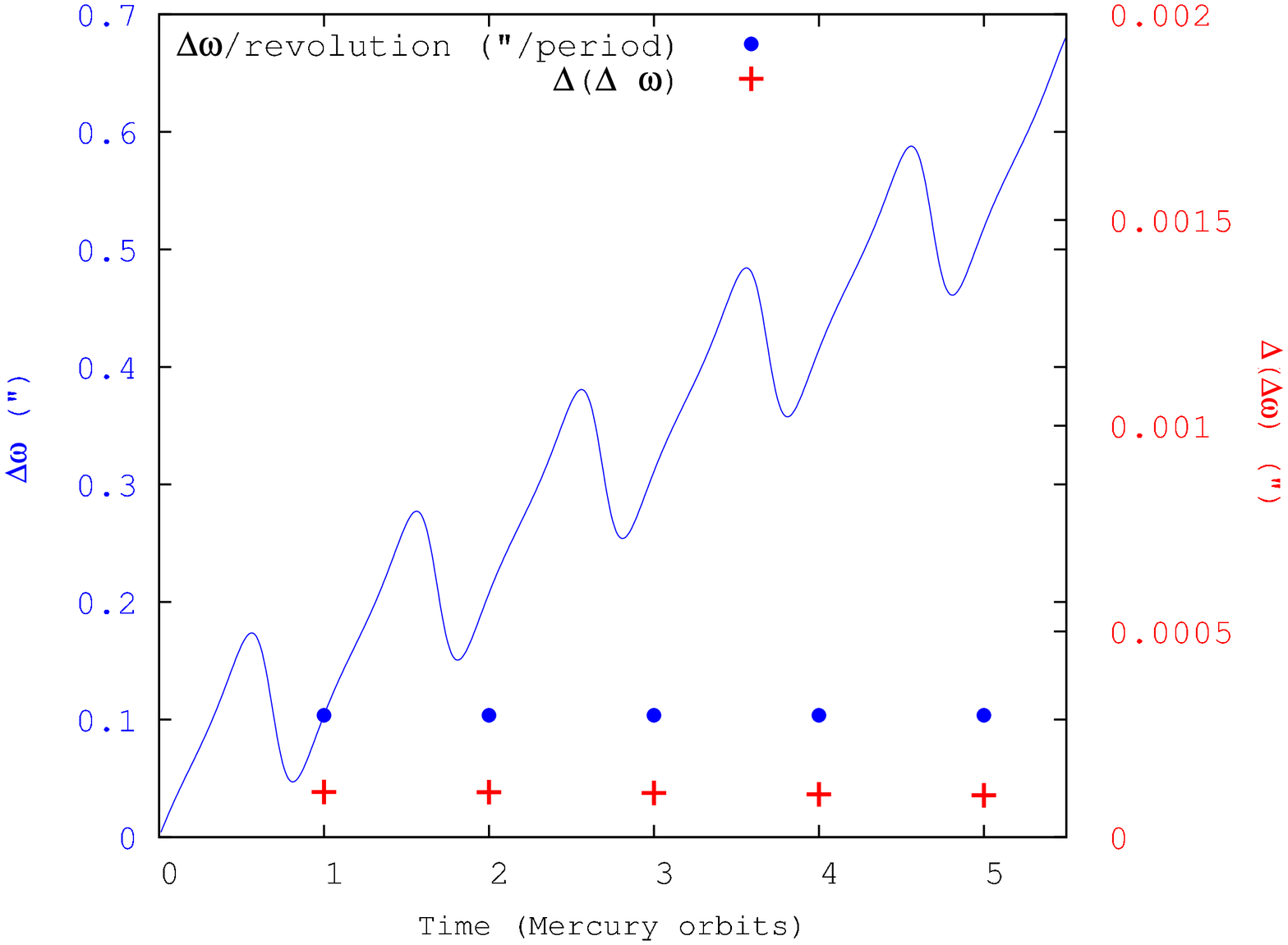}}
\caption{relativistic effect on Mercury's perihelion calculated with Lie integrator in arcsec (\textbf{\textcolor{blue}{-}}). $\Delta(\Delta \omega)$
represents the absolute difference between the 
theory and Lie value in arcsec (\textcolor{red}{$+$}). We also represented the perihelion precession per revolution
(\textcolor{blue}{$\bullet$}).} 
 \label{test_mercury}
\end{figure}

We also tested our algorithm by calculating the perihelion precession of Mercury and comparing with the expected value. The General Relativity 
predicts a secular precession of Mercury's perihelion which expression is \citep{balogh02}:

\begin{eqnarray}
\label{peri}
 \frac{d \omega}{dt} = \frac{6\,\pi\,GM_{\scriptscriptstyle \odot}}{c^{\scriptscriptstyle 2}\,a(1-e^{\scriptscriptstyle 2})} \approx
~~\mbox{0.103526}\, ''/\mbox{revolution}
\end{eqnarray}

\noindent
where $\omega$, \textit{a} and \textit{e} are respectively, the perihelion, semimajor axis and eccentricity of Mercury, \textit{M}${\scriptscriptstyle
\odot}$ the mass of the Sun.\\
Figure \ref{test_mercury} represents the relativistic effect on orbital parameter $\omega$ for 5 Mercury orbital periods. This graphic shows
that Mercury's 
perihelion position shifts of about 0.1 arcsec for each Mercury year. Moreover, we also compute the absolute difference $\Delta (\Delta\omega)$
between $\Delta \omega$ calculated 
with Eq. (\ref{peri}) and with the Lie integrator. This difference is about 1.10$^{\scriptscriptstyle -4}$ arcsec/revolution what shows the Lie value
is 
in a good agreement with the analytical result.

\subsection{Validation test for the Yarkovsky effect algorithm}

In this section, we test our Yarkovsky effect algorithm by calculating the secular drift ($da/dt$) of the semimajor axis caused by this effect. 
From the knowledge of $da/dt$, we can deduce a value for the constant component A$_{\scriptscriptstyle 2}$ in Eq. (\ref{eq_yarko}). 
As a matter of fact, the expression of the semimajor axis variations due to Yarkovsky effect along the transverse component, given by Gauss equations,
is:
\begin{eqnarray*} 
\frac{da}{dt} = \frac{2}{\sqrt{(1-e^2)}\,n}\,(1+e\,\cos\,\theta)\,\|\pmb{\gamma}_{\scriptscriptstyle Y}\|
\end{eqnarray*}

with $\theta$ the true anomaly. Thus, we deduce that the secular drift of the semimajor axis per revolution of the object is:\\
\begin{eqnarray}
 \Delta\,a = \frac{4\,\pi\,a}{k^{\scriptscriptstyle 2}\,(1-e^{2})}\,A_{\scriptscriptstyle 2}
\end{eqnarray}

\begin{table}[h!]
 \begin{center}
  \caption{\emph {Secular drift of semimajor axis per revolution calculated with Lie and Radau integrator.}}
  \label{yarko}
  \begin{tabular}{|c|c|c|c|c|c|c|}
   \hline
   \hline
   Asteroids  & Period & $A_{\scriptscriptstyle 2}$                & $da/dt$ (Chesley) & $da/dt$ (Radau) & $da/dt$ (Lie) \\
              & (day) &(AU/day$^{\scriptscriptstyle 2}$)         & km/rev & km/rev & km/rev \\
  \hline
   Golevka    & 1442.31 &-1.47$\times$10$^{\scriptscriptstyle -14}$& -0.3785 & -0.3616 & -0.3604 \\
   Apollo     & 651.17  &-4.67$\times$10$^{\scriptscriptstyle -15}$& -0.0642 & -0.0612 & -0.0611 \\
   Ra-Shalom  & 277.26  &-1.20$\times$10$^{\scriptscriptstyle -14}$& -0.0807 & -0.0781 & -0.0782 \\
   Bacchus    & 408.90  &-2.22$\times$10$^{\scriptscriptstyle -14}$& -0.1778 & -0.1675 & -0.1675 \\
   YORP       & 368.43  &-5.49$\times$10$^{\scriptscriptstyle -14}$& -0.3801 & -0.3595 & -0.3594 \\
   Hathor     & 283.32  &-2.35$\times$10$^{\scriptscriptstyle -14}$& -0.1622 & -0.1405 & -0.1405 \\
   Cerberus   & 409.94  &-1.46$\times$10$^{\scriptscriptstyle -14}$& -0.1313 & -0.1324 & -0.1323 \\
   Geographos & 507.70  &-2.68$\times$10$^{\scriptscriptstyle -15}$& -0.0246 & -0.0242 & -0.0242 \\
   Toro       & 583.92  &-1.13$\times$10$^{\scriptscriptstyle -15}$& -0.0125 & -0.0118 & -0.0117 \\
\hline
\hline
  \end{tabular}
 \end{center}
\end{table}

We propagated the motion of nine candidate asteroids for Yarkovsky perturbation over five Keplerian revolutions ($\Delta T$) around the Sun. The
asteroids considered here are taken from \cite{chesley08}, with the correspondent value of the da/dt given.
We compared the mean value of $da/dt$ calculated with Lie integrator, with Radau integrator to the one provided in \cite{chesley08}. The results
listed in table \ref{yarko} show that the value of $da/dt$ found with Lie integrator is in good agreement with Radau's one. Moreover, those values 
are not far from Chesley's one ($\sim 4\%$ or $5\%$ higher or lower). This difference can come from the fact that the Yarkovsky model used in our
simulation may be different than Chesley's one.

\section{Conclusion and Perspectives}
\label{sec:conclusion}

We presented in this paper a redesigned Lie integrator including relativistic and a simplified Yarkovsky forces. In addition to the CPU and accuracy
tests already performed, the numerical tests presented here show the great capacities of this extended Lie integrator that make it a useful and
complete integrator. 
This integrator can be useful when dealing with long-time integration and it can also handle simultaneous integration of massless bodies as well as 
Radau integrator. Besides, the integration of PHAs and the study of close encounters and impact probabilities can also be done now thanks to the 
relativistic acceleration and Yarkovsky effect algorithms.\\
For the future, the Lie integrator can be extended to the problem of the integration of comets's motion. The Yarkovsky effect algorithm is already 
a first approach for this study in that, the transverse vector is already implemented. Besides, one can consider to generalise the Lie integrator for
any position and velocity dependent forces.

\bibliographystyle{plainnat}
\bibliography{biblio}

%
%

\end{document}